\documentstyle[12pt,aaspp4]{article}
\def\et{{\it et al.}}

\def\del{\partial}
\def\<{{\langle}}
\def\>{{\rangle}}

\def\tW{{\tilde{\Omega}}}
\def\sA{{\mathcal{A}}}

\def\sD{{\mathcal{D}}}
\def\sE{{\mathcal{E}}}
\def\sL{{\mathcal{L}}}

\def\W{{\Omega}}

\def\sles{\lower2pt\hbox{$\buildrel {\scriptstyle <}
   \over {\scriptstyle\sim}$}}
\def\sgreat{\lower2pt\hbox{$\buildrel {\scriptstyle >}
   \over {\scriptstyle\sim}$}}
\lefthead{Popham and Gammie}
\righthead{Accretion in Kerr Metric}
\begin{document}
\title{Advection Dominated Accretion Flows in the Kerr Metric:
	II. Steady State Global Solutions}
\author{Robert Popham\altaffilmark{1} and Charles F. Gammie\altaffilmark{2,3} }
\affil{Harvard-Smithsonian Center for Astrophysics, MS-51 \\
60 Garden St., Cambridge, MA 02138}

\altaffiltext{1}{also Max-Planck-Institut f\"ur Astrophysik,
	Karl-Schwarzschild-Strasse 1, 85740 Garching, Germany}
\altaffiltext{2}{also Institute of Astronomy, Madingley Road,
	Cambridge CB3 0HA, United Kingdom}
\altaffiltext{3}{also Isaac Newton Institute, 20 Clarkson Road,
	Cambridge CB3 0EH, United Kingdom}

\today

\begin{abstract}

In a previous paper we have written down equations describing
steady-state, optically thin, advection-dominated accretion onto a Kerr
black hole (\cite{GP}, hereafter Paper I).  In this paper we survey the
numerical solutions to these equations.  We find that the temperature
and density of the gas in the inner part of the accretion flow depend
strongly on the black hole spin parameter $a$.  The rate of angular
momentum accretion is also shown to depend on $a$; for $a$ greater than
an equilibrium spin parameter $a_{eq}$ the black hole is de-spun by the
accretion flow.  We also investigate the dependence of the flow on the
angular momentum transport efficiency $\alpha$, the advected fraction
of the dissipated energy $f$, and the adiabatic index $\gamma$.  We
find solutions for $-1 < a < 1$, $10^{-4} \le \alpha \le 0.44$, $0.01
\le f \le 1$, and $4/3 < \gamma < 5/3$.  For low values of $\alpha$ and
$f$ the inner part of the flow exhibits a pressure maximum and appears
similar to equilibrium thick disk solutions.

\end{abstract}

\section{Introduction}

In advection-dominated accretion flows (ADAFs) the accreting gas flows
inward much more rapidly than it can cool.  The energy released by
accretion goes into heating the gas.  This is in marked contrast to
the usual thin accretion disk, where the radial velocity is small and
the accretion energy is efficiently radiated away.  If a black hole
accretes via an ADAF much of the accretion energy can be carried
across the horizon with the heated gas, reducing the luminosity well
below that of a comparable thin disk.  Paper I briefly summarizes the
development of advection-dominated disk theory.  Over the past few
years, this theory has had notable success in reproducing the observed
spectra of black hole candidate systems.  For a review of the theory
and applications of advection-dominated disk theory, see \cite{n97}.

Early models of advection-dominated flows around black holes
(\cite{ch96}, \cite{nak96}, \cite{nak97}, \cite{cal97}, \cite{nkh})
did not include a proper treatment of relativistic effects, but
instead used an approximate pseudo-Newtonian potential due to
\cite{pw80}.  Close to the black hole event horizon, the gas
temperatures and velocities can become extremely high.  This hot,
rapidly rotating and infalling gas should produce important observable
effects in the high-energy spectra of black hole candidates.
Relativistic effects not accounted for by the pseudo-Newtonian
potential are dominant in this innermost region.  The character of the
inner portion of the disk also depends strongly on the black hole spin
$a$, which is not included in the pseudo-Newtonian treatment.  For
these reasons, it seems clear that a fully relativistic model is
needed.

Recently, advection-dominated disk models have begun to include the
effects of general relativity, allowing a more accurate examination of
the flow close to the event horizon.  Both \cite{acgl} (AGCL) and
\cite{pa97} (PA) have presented disk solutions in the Kerr metric.  In
Paper I, we wrote down a set of disk equations in the Kerr metric, and
showed a few example solutions.  Our relativistic disk equations differ
in a number of respects from those of ACGL and PA.  These differences
are described in detail in Paper I.  The main differences with ACGL are
that we use a causal stress and include the relativistic enthalpy.  PA
also include these effects, although they use a more simplified
prescription to enforce causality.  Unlike PA, who use a polytropic
equation of state, we solve the energy equation assuming that a
constant fraction of the dissipated energy gets advected with the gas.
Also, unlike ACGL and PA, we use the height prescription of
\cite{alp}.

In this paper, we examine the structure of our numerical solutions for
steady state flow onto a Kerr black hole in detail.  In particular, we
explore the effects of changing the dimensionless black hole spin $a$,
the viscosity parameter $\alpha$, the advected fraction of the
dissipated energy $f$, and the adiabatic index $\gamma_0$ (ACGL have
shown solutions for three values of $a$, while PA presented a more
extensive set of solutions for various values of $a$ and $\alpha$).
We show that there is an equilibrium spin rate for black holes
accreting via ADAFs, above which the black hole will be spun down by
accretion.  We also show that there is a pressure maximum in the
vicinity of the last stable orbit for low $\alpha$ solutions
(cf. NKH).  Finally we show that the Bernoulli parameter can be
positive in some regions of the flow, which at least suggests the
possibility of a pressure-driven outflow.

To familiarize readers with our notation, we summarize our
relativistic disk equations and method of solution in \S 2.  In \S 3
we examine the dependence of our disk solutions on $a$, $\alpha$, $f$,
and $\gamma_0$.  \S 4 demonstrates that there is an equilibrium spin
rate for black holes accreting from ADAFs.  We discuss the
implications of our results in \S 5.

\section{Disk Equations and Solution Method}

\subsection{Summary of Disk Equations}

We use a relativistic version of the slim disk equations first
introduced by \cite{pb81} and \cite{mp82} to study the structure of
advection-dominated accretion disks.  The equations are presented in
detail in Paper I, but for the reader's convenience we summarize them
here.  The units are such that $G = M = c = 1$, where $M$ is the mass
of the black hole.

We begin with the continuity equation:
\begin{equation}
4\pi r^2 \rho H_\theta V \left(\sD\over{1 - V^2}\right)^{1/2}
	= -\dot{M}.
\end{equation}
Here $r$ is the Boyer-Lindquist radius, $\rho$ is the rest mass
density, $H_\theta \equiv H/r$ is the relative disk thickness, $V$ is
the radial velocity measured in a corotating frame, $\sD \equiv 1 -
2/r + a^2/r^2$ is a relativistic correction factor, and $\dot M$ is
the rest mass accretion rate.

The gas energy equation is
\begin{equation}\label{ENCONS}
V \left(\sD\over{1 - V^2}\right)^{1/2}
\left({\del u\over{\del T}} {d T\over{d r}} - 
{p\over{\rho}}{d\rho\over{d r}}\right) = f\Phi.
\end{equation}
Here $u$ is the internal energy per unit proper volume, $T$ is a
dimensionless temperature such that the pressure $P = \rho T$, 
$\Phi$ is the dissipation function, and $f \le 1$ is a parameter used to
mock up the effects of cooling, if any.  The dissipation function
$\Phi$ is given in Paper I (eq. (68) and Appendix A).  We also take
\begin{equation}
u = \rho T g(T) \equiv \rho T \left({4/(\gamma_0 - 1) + 15 T\over{
	4 + 5 T}}\right),
\end{equation}
which is a good approximation to the exact relativistic equation of
state for an ideal relativistic Boltzmann gas (see \cite{ch39}) when
$\gamma_0 = 5/3$.  Because the accretion flow is assumed to consist of
a mixture of magnetic fields and ionized plasma, $\gamma_0$ can be
smaller than $5/3$.

The radial momentum equation is
\begin{equation}\label{RMCONS}
{V\over{1 - V^2}} {d V\over{d r}} = f_r - {1\over{\rho\eta}}
	{d p\over{d r}},
\end{equation}
where
\begin{equation}
f_r \equiv -{1\over{r^2}}{\sA \gamma_\phi^2\over{\sD}}
	(1 - {\Omega\over{\Omega}}_+) (1 - {\Omega\over{\Omega}}_-).
\end{equation}
The $f_r$ term combines the effects of gravity and rotation,
where $\sA \equiv 1 + a^2/r^2 + 2 a^2/r^3$ and $\gamma_\phi^2 = 1 + l^2
(1-V^2) / (r^2 \sA)$, $\Omega = u^\phi/u^t$ is the angular velocity,
and $\Omega_\pm = \pm (r^{3/2} \pm a)^{-1}$.  The radial acceleration,
on the left-hand side, is given by the difference between $f_r$ and the
pressure gradient force, where $\eta$ is the relativistic enthalpy
$\eta \equiv (\rho + p + u)/\rho$.

The angular momentum conservation equation is
\begin{equation}\label{AMCONS}
\dot{M} l\eta - 4\pi H_\theta r^2 {t^r}_\phi = \dot{M} j.
\end{equation}
Here $l$ is the specific angular momentum of the accreting gas, $j =
const.$ is the angular momentum accretion rate per unit rest mass
accreted.  The remaining term gives the viscous angular momentum
transport rate, where $t^r_{\phi}$ is the viscous stress.  Notice that
we solve for $j$ as an eigenvalue, as do NKH and ACGL.  PA prespecify
$j$ (which they label $L_0$) and iterate in $j$ when they want to find
a solution with a specified value of $l$ at the outer edge of the
disk.  \cite{ch96} also prespecifies $j$.

The calculation of $t^r_{\phi}$ is rather lengthy, and the reader is
referred to \S 4 of Paper I for a full mathematical discussion.  To
briefly summarize the relevant physics, we begin by calculating the
shear rate $\sigma$ in the local rest frame of the accreting gas.  This
gives a complicated expression which is closely approximated by the
thin disk shear rate derived by \cite{nt73}, $\sigma_{thin} = (1/2) \sA
\gamma_\phi^2 r d\Omega/dr$ (see the Appendix of Paper I for a
discussion).  Next we specify the relation between the turbulent shear
stress and the shear rate (rate of strain).  The simplest prescription
is the usual Navier-Stokes form in which stress is simply proportional
to rate of strain.  As is well known, the resulting system of equations
is acausal.  To preserve causality, we use a relativistic version of a
prescription that originated with \cite{max67} and Cattaneo (1948,
1958) and was discussed in the context of accretion disk theory by
\cite{ps94} (\cite{nlk} give a mathematically equivalent formulation).
This modification limits the propagation of viscous effects to a finite
speed $c_\nu$, which we take to be $c_\nu = \sqrt{\alpha} c_s$ (our
solutions are not sensitive to the precise value of $c_\nu$).  The
viscous stress $t^r_\phi$ is then obtained by transforming back from
the local rest frame to the Boyer-Lindquist frame.

The equation of vertical mechanical equilibrium is
\begin{equation}\label{VERT}
H_\theta^2 = {p\over{\rho\eta r^2 \nu_z^2}},
\end{equation}
where $\nu_z$ is an effective vertical frequency.  We adopt the
expression derived by \cite{alp} for $\nu_z$:
\begin{equation}
\nu_z^2 = {l^2 - a^2 (\sE^2 - 1)\over{r^4}},
\end{equation}
where $\sE = -u_t$ is the ``energy at infinity'', which is conserved 
along geodesics. 

Finally, the inward flux of mass-energy is 
\begin{equation}\label{EDOT}
\dot{E} = 4\pi H_\theta r^2 [-(\rho + u + p) \sE u^r + {t_t}^r],
\end{equation}
where $t$ is the viscous stress tensor.  We do not use this equation to
find our solution.  But when $f = 1$, $\dot{E}$ is constant with $r$,
and $\dot{E} \simeq \dot{M}$,  because $T \ll 1$ at large $r$.  In this
case equation (\ref{EDOT}) is a check on our numerical accuracy.   It
is generally satisfied to better than one part in $10^{-3}$.  For $f <
1$, however, $\dot{E}$ is not conserved, since some of the rest-mass is
radiated away.  Recall that $\dot{E}$ at the horizon, and not
$\dot{M}$, is the true rate of change of the mass of the black hole.

\subsection{Critical Points and Boundary Conditions}

We set the outer edge of the ADAF at $10^4$ Schwarzschild radii, or $r
= 2 \times 10^4 \, G M/c^2$, and the inner edge just outside the event
horizon at $r = (1 + \sqrt{1 - a^2}) \, G M/c^2$, where $\sD = 0$.  

At the outer edge we impose two boundary conditions: $\Omega$ and $c_s$
must equal their values in the self-similar advection-dominated
solution of \cite{ny94}.  We have also obtained solutions with the thin
disk values for $\Omega$ and $c_s$ at the outer edge, but our solutions
adjust to the self-similar profile within a short distance of the
boundary.  NKH found a similar result, since their equations are
essentially equivalent to ours in the non-relativistic limit.  Thus the
self-similar solution is the ``natural'' state of the flow far from the
black hole, and is an appropriate outer boundary condition.

Two other conditions on the flow are provided by the requirement that
the flow pass smoothly through two critical points.  The first is the
sonic point $r_s$, where $|V| \simeq c_s$.  The second is the
``viscous point'' $r_v$ associated with the finite propagation speed
of viscous effects, where $|V| \simeq c_\nu$.  Associated with each
critical point are two conditions that must be satisfied for a smooth
flow, as well as one degree of freedom, the location of the critical
point itself.

The final boundary condition normalizes the density (the density
appears in the basic equations only in the form $d\ln\rho/d r$).  For
simplicity we set the normalization constant so that $\dot{M} = 1$.  We
now have all the boundary conditions required to solve the four
first-order ordinary differential equations for $V,l,\rho,$ and $T$,
and to find the eigenvalue $j$.  In particular, consistent with
causality, no boundary conditions are applied at the event horizon.

\subsection{Method of Solution}

We solve the system of equations listed above using a relaxation
method.  We solve for the structure of the flow in three radial
regions: the subsonic, ``sub-viscous'' outer zone, where $r > r_v$, the
subsonic, ``super-viscous'' middle zone, where $r_s < r < r_v$, and the
supersonic, ``super-viscous'' inner zone, where $r < r_s$.  We first
solve the outer zone for a specified value of $r_v$, and then solve the
middle zone for the same $r_v$.  We compare the outer and middle zone
solutions to see whether they match up at $r_v$; if not, we change
$r_v$ and repeat the procedure until the outer and middle solutions
match up.  Then, using the variable values from the middle solution at
$r_s$ as boundary conditions, we solve for the inner zone.

As is well known, our independent variable, the Boyer-Lindquist $r$, is
ill-behaved near the horizon when $|a| \to 1$.  This might raise some
concerns about the numerical accuracy of our solutions.  We have taken
several steps to control for this, however.  First, we use a fixed
number of grid points in each radial region.  As $a \to 1$, the sonic
radius moves sharply inward, and this greatly increases our numerical
resolution close to the horizon.  Second, we have convergence-tested
our solutions by increasing the number of grid points.  No significant
changes were found at increased resolution.  Finally, notice that
conditions close to the horizon do not influence the solutions at
larger radius because our fundamental equations are causal:  the
solution in the inner region is obtained by integrating inward from the
sonic point.  In fact, many of the important parameters of the flow,
such as the angular momentum accretion rate $j$, are fixed by
conditions at and outside the sonic point.

\section{A Survey of Solutions}

Here we describe the changes in our solutions that result from changes
in four basic parameters: the black hole spin $a$, the viscosity
parameter $\alpha$, the advected fraction of the dissipated energy $f$,
and the adiabatic index $\gamma$.  We begin from a base solution with
$a=0$, $\alpha = 0.1$, $f=1$, and $\gamma = 1.4444$ (which corresponds
to equipartition between gas and magnetic pressure).  This solution was
described in detail in Paper I, and it appears in each of the sets of
solutions described below.  Recall that physical units may be recovered
as follows: radial velocity is $V c$, angular momentum is $l G M/c$,
density is $\rho\dot{M} G/c^3$ (since the mass accretion rate is set to
1), and temperature is $T\bar{m} c^2/k$, where $\bar{m}$ is the mean
molecular weight and $k$ is Boltzmann's constant.

It is worth pointing out immediately that the marginally stable orbit
plays no special role in many of our solutions.  While the sonic point
$r_s$ often lies close to the radius of the marginally stable orbit,
there is no abrupt change in the character of the flow there.  This is
because pressure gradients play an important role in the radial
structure of ADAFs; unlike thin disks, ADAFs are not in centrifugal
balance.

\subsection{Black Hole Spin $a$}

We begin by examining the dependence of the accretion flow on the black
hole spin parameter $a$.  We solve for values in the range $-1 < a <
1$, allowing for the possibility of a counterrotating hole.  The
horizon lies at $r = 1 + (1-a^2)^{1/2}$, i.e. at $r = 2$ for $a = 0$
and at $r = 1$ for $a = \pm 1$.  The resulting profiles of $\rho$, $V$,
$\W$, $l$, $T$, and $H/R$ are shown in Figure 1 for solutions with
$a=-0.999$, -0.9, -0.5, 0, 0.5, 0.9, and 0.999.

Notice that changes in $a$ have little effect far from the hole.  Also,
notice that in general the positive-$a$ solutions differ from the $a=0$
solution more than the corresponding negative-$a$ solutions.  This is
especially dramatic in $\rho$ and $T$: the $a=0.999$ solution reaches
very high density and temperature near the horizon, while the
$a=-0.999$ solution has $\rho$ and $T$ at the horizon similar to the
$a=0$ solution.  

The positive and negative-$a$ solutions also show very different $V$
profiles close to the hole.  The positive-$a$ solutions accelerate
rapidly and plunge into the hole, while the negative-$a$ solutions
accelerate rapidly farther from the hole and only gradually near the
hole, coasting across the horizon.

Figure 1 also shows the positions of the sonic and viscous points for
each solution.  The sonic and viscous points move to smaller $r$ as
$a$ increases, as expected since the ``plunge'' toward the hole occurs
at smaller radii.  At $a=0$, the sonic point is at $r \simeq 6.41$,
close to the last stable orbit.  At $a=0.999$, it has moved in to $r
\simeq 1.94$, while for $a=-0.999$ it moves out to $r \simeq 9.20$.
The viscous point always lies much farther from the hole; in general
$r_v \sim 4-5 r_s$.  This is due to our setting $c_\nu^2 = \alpha
c_s^2$, so that for $\alpha = 0.1$ we have $c_\nu \sim 0.3 c_s$.

The value of $\W$ at the horizon should be equal to $\omega \equiv 2 a
/ (r^3 + a^2 r + 2 a^2)$ (denoted by a dashed line in Fig. 1) due to
the effects of frame-dragging.  This results in the symmetric
distribution of inner $\W$ values between -0.5 and 0.5 seen in Fig. 1.
For all but the high-$a$ solutions, $\W$ shows a boundary layer-like
profile, reaching a peak and then dropping rapidly to $\omega$ at the
horizon.  Had we plotted $\tW \equiv \W - \omega$, the profiles for
all values of $a$ would resemble the $\W$ profile for $a=0$.

The specific angular momentum $l \eta$ drops steadily as $a$
increases. Notice that $l \eta$ includes the angular momentum associated
with both the mass density and the energy density of the gas.  The
inner value of $l \eta$ decreases from over 3 to less than 2 as $a$
goes from -0.999 to 0.999.  The actual rate of angular momentum
accretion by the hole per unit rest mass accreted is $j$.
Values for $j$ are generally close to the horizon
values of $l \eta$.  This means that the rate of viscous angular
momentum transfer is very small near the horizon; however, it is not
zero.

The relative disk thickness $H/r$ also varies dramatically with $a$.
Notice that for all values of $a$, $H/r$ is smaller near the horizon,
then increases to $\sim 1$ at large radii.  Negative-$a$ flows are
much thinner than positive-$a$ flows.

\subsection{Viscosity Parameter $\alpha$}

Figure 2 shows a set of solutions with $\alpha = 0.001$, 0.003, 0.01,
0.03, 0.1, 0.3, all with $a = 0$, $f = 1$, and $\gamma = 1.4444$.
Both the sonic and viscous points move steadily outward as $\alpha$
increases, since a larger $\alpha$ gives a larger viscosity
coefficient for given $c_s$ and $H$, producing larger radial
velocities $V$.  This removes more angular momentum and dissipates
more energy, resulting in smaller values of $l \eta$ and higher
temperatures for larger $\alpha$.  The higher temperature gives a
larger relative disk thickness $H/r$, and this combined with the
larger radial velocity makes the density decrease as $\alpha$
increases.

The density and temperature profiles show an interesting effect at low
$\alpha$: the development of a local maximum in density and temperature
just outside the last stable orbit.  This effect was also noted by NKH
and CAL.  The maximum becomes more pronounced as $\alpha$ decreases.
It is faintly seen as an inflection in the density profile at $\alpha =
0.01$, but by $\alpha = 0.001$ it produces a very large, wide peak with
a density 50\% larger than the density at the horizon.  The maximum in
the temperature profile is less dramatic, but the combined effect of
the two is to produce a pressure maximum which affects the dynamics of
the flow.  On the inner side of this pressure maximum, the pressure
gradient force points inward.  This inward force is balanced in part by
more rapid rotation: $\W$ increases as $\alpha$ decreases.  In fact the
$\alpha = 0.001$ solution even has a super-``Keplerian'' rotation in a
small region extending from $r \simeq 5-8$.

\subsection{Advected Fraction $f$}

The parameter $f$ was introduced to mock up the effects of cooling.
The case of greatest interest for modeling ADAFs is when cooling
is unimportant, so $f \simeq 1$.  We can reduce $f$ so that it is
much less than $1$, however, and in this case we should approximately
recover a thin disk, albeit with a radial temperature structure that
is not consistent with any realistic cooling function.

Figure 3 shows solutions for $f = 1$, 0.3, 0.1, 0.03, 0.01.  All of the
solutions have $a=0$, $\alpha = 0.1$, and $\gamma = 1.4444$. The
dramatically lower temperatures in low-$f$ solutions reduce $c_s$ and
$H$ and thus reduce the viscous stress.  As a result, this sequence of
solutions strongly resembles the $\alpha$ sequence shown in Fig. 2.
For small values of $f$, as for small values of $\alpha$, a maximum
develops in density, temperature, and pressure in the inner disk for
small values of $f$, resulting in super-Keplerian values of $\W$.  The
relative disk height drops as $f$ decreases, and the removal of angular
momentum is less efficient.  The major difference between the two
sequences is that low values of $f$ produce much lower temperatures and
higher densities than low values of $\alpha$.

The low-$f$ solutions are indeed close to a thin disk.  In particular,
the sonic point occurs close to the last stable circular orbit (this is
true for solutions with a variety of $a$ at low $f$).  Also, the
energy-at-infinity $\sE$ and angular momentum $l$ are close to their
circular orbit values.  Of course, this is expected due to the low
temperatures of these models, so that radial pressure gradients play a
negligible role.  Nevertheless, it is good to see this expectation
confirmed.

\subsection{Adiabatic Index $\gamma$}

The adiabatic index is not fixed by the temperature because some
(unknown) fraction of the total pressure is contributed by the magnetic
field, while the rest is contributed by gas pressure.  Figure 4 shows a
sequence of solutions for a range of values of $\gamma = 1.3333$,
1.4444, 1.55, 1.66.  These correspond to the fraction of the pressure
contributed by the gas pressure $\beta = 0$, 0.5, 0.8, 0.99,
respectively.  All solutions have $a=0$, $\alpha = 0.1$, and $f = 1$.

The high-$\gamma$ solutions get much hotter than the
low-$\gamma$ ones, increasing the effective viscosity coefficient.
This produces higher velocities, larger relative disk heights, and
removes more angular momentum.  As $\gamma$ approaches $5/3$, these
effects become extreme.  For example, the $\gamma = 1.66$ solution
has unphysically large values of $H/R$.

\section{Equilibrium Black Hole Spin Rates for Advection-Dominated Accretion}

The black hole accretes both mass and angular momentum, so the
dimensionless black hole spin parameter $a \equiv J c/G M^2$ ($J$ is
the black hole angular momentum) changes with time.  In dimensionless
form,
\begin{equation}\label{DADT}
{d a \over{d t}} = \dot{M} j - 2 a \dot{E},
\end{equation}
where $\dot{E}$ is the inward flux of mass-energy.  Thus if $j = 2
a \dot E / \dot M$, then $d a/d t$ vanishes and the black hole has
reached an equilibrium spin $a_{eq}$.  The value of $a_{eq}$, if it
exists, will depend on the input parameters $a$, $\alpha$, $f$, and
$\gamma$.  Notice that when $f = 1$, as in most of the solutions
presented here, $\dot{E} = const. \simeq \dot{M}$, and $a_{eq}$ is
reached when $j = 2a$.

The equilibrium spin is of considerable astrophysical interest since
the observable properties of accretion flows change sharply in the
neighborhood of $a = 1$.  In an influential paper, \cite{b70} showed
that a black hole accreting from a thin disk reaches $a = 1$ in a
finite time.  \cite{th74} pointed out that if one takes account of the
preferential accretion of angular momentum by the hole, then $a_{eq} =
0.998$.  \cite{al80} suggested that even larger values of $a_{eq}$
might still be obtained from a thick disk, based on numerical
calculations that showed that the inner edge of these disks was located
somewhere between the last stable orbit and the last bound orbit.

In order to calculate $a_{eq}$, we have obtained sequences of
solutions with increasing $a$.  As $a$ increases, $j$ decreases, and
when $j = 2a \dot E / \dot M$ we have found $a_{eq}$.  In Figure 5
above we show the dependence of $j$ on $a$ for the six values of
$\alpha = 0.001$, 0.003, 0.01, 0.03, 0.1, 0.3, for which $a=0$
solutions were shown in Fig. 1.  In all cases, $j$ decreases as $a$
increases, and decreases faster as $a$ approaches 1, particularly in
the low-$\alpha$ solutions.  Here $f=1$, so $a_{eq}$ is reached when
$j=2a$, as denoted by the dotted line.  Notice that $a_{eq}$ is quite
far from $a=1$ for high-$\alpha$ flows: for $\alpha=0.3$, we have
$a_{eq} \simeq 0.8$.  On the other hand, for low-$\alpha$ flows,
$a_{eq}$ is very close to 1: for $\alpha = 0.001$, we have $a_{eq} =
0.999965$.  The variation of $a_{eq}$ with $\alpha$ is shown in Figure
6.  For the sequence of $\alpha = 0.001$, 0.003, 0.01, 0.03, 0.1, 0.3,
we find $a_{eq} = 0.999965, 0.99972, 0.9975, 0.985, 0.930, 0.806$.
For $\alpha \leq 0.03$ we find the approximate relation $1 - a_{eq}
\simeq 10 \alpha^{1.8}$.

We also obtained sequences in $a$ for values of $f$ between 0.01 and 1,
using $\alpha = 0.1, \gamma = 1.4444$.  For $f < 1$ we no longer have
$\dot E \simeq \dot M$, so the full form of equation (\ref{DADT}) is
required.  We find $a_{eq} = 0.930, 0.988, 0.9984, 0.99983, 0.999979$,
for $f = 1, 0.3, 0.1, 0.03, 0.01$, respectively, as shown in Figure 6.
Here again we find an approximate power-law relationship $1 - a_{eq}
\simeq 0.1 f^{1.8}$.

These results have some interesting implications.  If most of the mass
of a black hole is accreted through an ADAF, then the hole cannot
approach $a = 1$.  All advection-dominated models of observed systems
that have appeared in the literature have high values of $\alpha \sim
0.3$, since solutions with low $\alpha$ have extremely low
luminosities.  According to the results given above, the black holes in
these systems will not be able to spin up past $a \sim 0.8$.  The most
dramatic effects of the black hole spin on the flow, which appear when
$a$ nears unity, will not be seen in such systems.

Of course, the accretion of photon angular momentum is not included in
our calculation, but this is justified where $1 - f$ is small.  In our
flows, small values of $f$ correspond to thin disks, which radiate a
fraction $1-f$ of the energy dissipated in the disk.  The limiting
spin $a_{eq}$ also depends on $f$, and we find that flows with $f
~\sgreat~ 0.1$ will have $a_{eq} < 0.998$, so they should stop spinning
up before reaching the photon-transport limit.  Flows with $f < 0.1$
have $a_{eq} > 0.998$, so here the photon-transport limit is relevant.

\section{Discussion}

\subsection{Comparison to Earlier Solutions}

Our disk equations differ in a number of respects from those used by
other authors, as detailed in Paper I and above.  Also, the solutions
described above cover a wider range of parameter space than previous
studies.  We have found solutions over the ranges of $a=-0.99999$ to
$0.99999$, $\alpha = 0.0001-0.44$, $f = 0.01-1$, and $\gamma =
1.3333-1.66$.

Our sequence of solutions for $\alpha = 0.001 - 0.3$ resembles the
solutions of NKH in many respects.  We find sonic radii ranging from
$r_s \simeq 4.28 - 9.41$ as $\alpha$ increases, similar to the range
$r_s = 4.20 - 10.63$ found by NKH.  Our solutions also develop
pressure maxima at small values of $\alpha$, as discussed below.
Since NKH used the pseudo-Newtonian potential, their solutions have
some unphysical properties near the horizon, which are avoided by our
use of the full relativistic equations.

Our relativistic solutions with $a=0$ produce emission spectra which
are qualitatively similar to those calculated from the NKH solutions.
The densities tend to be somewhat higher near the horizon in our
solutions due to the inclusion of relativistic effects, and this
increases the luminosity of the flow substantially.  This was shown by
\cite{n98}, who calculated the emission spectrum of Sgr A* using the
relativistic dynamical solutions described in this paper.  They were
thus able to make a direct comparison between the spectrum computed
from our solution and that computed from a pseudo-Newtonian solution
as described by NKH.  This showed that the inclusion of relativistic
dynamics increased the emission by a factor of 10-100 in the infrared.
Note that relativistic photon transport effects are not included in
these spectra, apart from including the gravitational redshift.
Also, the effects of using rotating black hole solutions have yet to be
explored; however, it is clear that the substantial increases in
temperature and density near the horizon in these solutions will have
a dramatic effect on the emission spectrum.

ACGL presented solutions with $a = 0$, 0.5, and 0.99 for $\alpha =
0.1$, which included bremsstrahlung cooling.  The shape of the $l(r)$
curves is similar to ours; however, the solutions differ in some other
respects.  First, the ACGL solutions have larger values of $l$ at the
inner edge of the flow. They find $l_{in} \simeq 3.2, 2.6, 1.7$ for $a
= 0, 0.5,0.99$, respectively, whereas our solutions have $l_{in} =
2.14, 1.79, 0.93$ for these same values of $a$.  One reason for this is
that their $l(r)$ curves only extend in to the Schwarzschild radius at
$r=2$, even for the $a=0.5$ and $a=0.99$ solutions, where the horizon
sits at smaller radii.  Also, in ACGL's scheme, the angular momentum
eigenvalue which we call $j$ is equal to $l_{in}$ (or in their
notation, $\sL = \sL_0$), whereas in our solutions, $j$ can differ
substantially from $l_{in}$ due to viscous torques.  Our values of $j$
for these solutions are 2.62, 2.31, and 1.73, which are somewhat closer
to the ACGL values.  Another difference between our solutions and those
of ACGL is that their solutions show a maximum in $\log c_s$ and
inflections in $\log P$ close to the inner edge, whereas our solutions
have $\log T$ and $\log P$ increasing smoothly all the way in to the
horizon.

PA have presented the most extensive survey of solutions to date.  They
find two types of viscous disk solutions: ``type I'' solutions at low
values of $\alpha$ ($\alpha = 0.001-0.045$ in the solutions shown) and
``type II'' solutions at high $\alpha$ ($\alpha \geq 0.3$ in the
solutions shown).  One major difference between PA's scheme and ours is
that they specify the angular momentum eigenvalue $j$ (which they call
$L_0$) while we solve for it.  Their type I solutions with $\alpha$
ranging from 0.01 to 0.045 all have the same value of $L_0$, and show
great variation in their angular momentum profiles at large $r$,
whereas our solutions tend to have rather similar values of $l$ at
large $r$.  PA's type I solutions also have a maximum in the sound
speed outside the sonic point, and the sound speed then decreases down
to the horizon.  This may be a consequence of their polytropic equation
of state, which requires that the sound speed must decrease inward if
the radial velocity increases inward more rapidly than $\sim r^{-1}$.
Their lowest-$\alpha$ solutions at $\alpha = 0.001 - 0.008$ have a
super-Keplerian region in the inner part of the disk.  In the type II
solutions shown by PA, the sonic point occurs far out in the flow at $r
\sim 30-60$.  This is a much larger sonic radius than in our
high-$\alpha$ solutions: our $\alpha = 0.3$ solution has $r_s = 9.41$.
According to PA, their type II solutions have the sonic point located
at an outer critical point rather than an inner critical point as the
Type I solutions do.

Unlike PA, who found two types of solutions, and \cite{ch96}, who has
found solutions where the flow goes through radial shocks, we find
smooth solutions for the entire range of parameter values, and we see
no evidence for sudden transitions between different types of
solutions.  In order to look for alternate solutions, we varied $r_v$
and solved for the outer and middle sections of our solutions to see
whether these sections would match up for an alternate value of $r_v$.
Despite looking for alternate solutions with several choices of
parameters, including low values of $\alpha$, we found no additional
solutions.

It is worth commenting in some detail on why our solutions are
shock-free while those of \cite{ch96} are not.  The origin of this
difference lies in how the boundary conditions are treated.  The
difference is most easily explained by analogy with the problem of
spherical (Bondi-Hoyle) accretion.  To solve the Bondi problem one
specifies the density and temperature, but not the radial velocity
(nor, equivalently, the accretion rate), at large distance from the
accreting object.  The velocity at large radius is adjusted until the
flow passes smoothly through the sonic point.  This velocity, or the
accretion rate, is thus an eigenvalue of the problem and must be
solved for self-consistently.  This approach has been validated by
numerically solving a realistic initial value problem and showing that
it settles down to the Bondi solution.  If one were to specify the
radial velocity or accretion rate the flow would not generally pass
smoothly through the sonic point; rather, it would shock.

Our treatment is analogous to the standard treatment of the Bondi
problem, except that in our case the mass accretion rate is specified,
while the angular momentum accretion rate $j$ is the eigenvalue.  It is
adjusted so that the flow passes smoothly through the sonic point.
Chakrabarti's approach, on the other hand, is analogous to prespecifying
the mass accretion rate in the Bondi problem: he fixes $j$.  As a result
the flow does not generally pass smoothly through the sonic point and
his solutions contain one or more shocks.  Our intuition, and the
analogy with the Bondi problem, suggest that this is not the correct
approach.  The issue can only be settled conclusively, however, by
solving a realistic initial value problem and showing that it converges
to one solution or the other.  This has not yet been done.

\subsection{Local Pressure Maxima}

Our low-$\alpha$ solutions have a maximum in pressure, density, and 
temperature in the inner disk.  Figure 7 shows density contours in the
inner parts of the flow for various values of $a$ and $\alpha$.  The
contour plots are in the $x = r\sin(\theta), z = r\cos(\theta)$
plane, and assume that $\rho \propto \exp(-\theta^2/(2 H_\theta^2))$.
These solutions resemble the low-$\alpha$ solutions of NKH.  They are
also similar to the thick disk models developed by a number of workers
(see \cite{fm76}, \cite{pw80}, \cite{rbbp}).  In particular they have 
a pressure maximum, approximately constant specific angular momentum,
and a region of super-Keplerian rotation where pressure support is
important.  Generally the pressure maximum lies close to the last stable
orbit, and the sonic point lies close to the marginally bound orbit.
Our approximations are most accurate for these low-$\alpha$ solutions in
that $H/R$ is relatively small (see Figure 2) and the low viscosity
implies a disk with relatively low turbulent velocities.

It is not entirely clear that the low-$\alpha$ models are relevant,
however.  A lower limit to the efficiency of angular momentum transport
is set by the existence of the global hydrodynamic instability of
\cite{pp84}.  Simulations of the nonlinear development of the
magnetorotational instability (\cite{bgh95}) suggest an even larger
lower limit of order $\alpha \sim 0.01$.  For $\alpha$ as large as this
a pressure maximum does not develop, and the pressure increases smoothly
down to the event horizon.

A new result of this study is the similarity of low-$f$ solutions to
low-$\alpha$ solutions.  Our $f=0.01$ and $f=0.03$ solutions show
pronounced density and pressure maxima in the same region of the flow,
just outside the last stable orbit.  They also have super-Keplerian
rotation in this region.  However, the low-$f$ solutions have the
sonic point at $r_s \simeq 5.72$ for $f=0.03$ and at $r_s \simeq 5.85$
for $f=0.01$, closer to the last stable orbit than the marginally
bound orbit.

\subsection{Bernoulli Parameter}

The Bernoulli parameter $Be$ measures the sum of the kinetic energy,
potential energy, and enthalpy of the gas.  Narayan \& Yi (1994, 1995)
pointed out that self-similar advection-dominated flows have a positive
Bernoulli parameter for $f > 1/3$.  The positivity of $Be$ suggests the
possibility of a pressure-driven outflow; $Be$ is conserved for
adiabatic, inviscid flows, so that gas with $Be > 0$ could flow outward
adiabatically and still have positive kinetic energy at large radius.

In its relativistic form, $Be = \eta \sE - 1$, where $\sE \equiv
-u_t$.  Figure 8 shows $Be$ for the solutions shown in Figs. 1--4,
illustrating the variation of $Be$ with radius and with $a$, $\alpha$,
$f$, and $\gamma$.  Solutions with small values of $a$ reach only small
positive values of $Be$.  In these solutions, $Be$ tends to peak at $r
\sim 6-20$, and reaches peak values of $Be < 0.01$ (see Fig. 8b,d).
Solutions with $a$ approaching unity can have substantially larger
values of $Be$ which peak at the horizon (Fig. 8a).  Solutions with $f
< 1$, which radiate away a fraction $1-f$ of the dissipated energy,
have $Be < 0$.  At large radii, the solution with $f=0.3$ has $Be$ only
slightly negative, as expected from the self-similar result that ADAFs
with $f > 1/3$ have $Be > 0$.

Variations in the Bernoulli parameter are directly related to the
radial viscous energy flux, since the total radial flux of mass energy
is conserved.  We have
\begin{equation}
Be = {\dot{E}\over{\dot{M}}} - 1 + {t_t^r\over{\rho u^r}}
\end{equation} 
where $t_{\mu\nu}$ is the viscous stress tensor.  Recall that $\dot{E}$
at the event horizon is the actual rate of change of the black hole
mass; furthermore, $\dot{E} = const. \simeq \dot{M}$ when $f = 1$.  One
can show that, if $S$ is the shear stress measured in the local rest
frame of the fluid, then
\begin{equation} \label{ENERFLUX}
t_t^r = -r\sD u^\phi S.
\end{equation}
Since $u^\phi \sim \sD^{-1}$ at the horizon, $t_t^r$ is finite at the
horizon.  Figure 9 shows the run of $-t_t^r$, the
{\it outward} viscous energy flux, for solutions with several values of
$a$ but otherwise with the standard parameters.

The existence of a finite outward angular momentum and energy flux at
the horizon suggests a violation of causality, yet our solution is
manifestly causal.  How can this be?  It turns out that these fluxes
appear because of how the flow is divided into a mean and fluctuating
part.  Consider a simple example: an accretion flow consisting of a
turbulent, unmagnetized fluid with negligible density fluctuations.  At
a given event, the flow has angular momentum $l + \delta l$ and radial
four-velocity $u^r + \delta u^r$.  The mean flow is defined so that
$\<\delta l\> = 0$ and $\<\delta u^r\> = 0$, where the brackets denote
an average over $t,\theta,\phi$.  Then the outward flux of angular
momentum is
\begin{equation}
{T_\phi}^r = (\rho + u + p)(l + \delta l)(u^r + \delta u^r),
\end{equation}
using the definition of the perfect fluid stress tensor.  Averaging,
\begin{equation}
{T_\phi}^r = (\rho + u + p)[l u^r + \<\delta l \delta u^r\>].
\end{equation}
The first term in brackets is due to the mean flow; the second term is
what we have called the ``viscous'' angular momentum flux.  Evidently
the outward flux of angular momentum (or energy; the same
considerations apply to ${T_t}^r$) appears because of how we have
divided the flow into a mean and fluctuating part.  Correlations in the
fluctuations merely bias the angular momentum of accreting fluid
elements.  Causality is preserved.

\subsection{Assumptions and Limitations}

Finally, it is worth offering a frank discussion of the assumptions
behind our solutions and their limitations.  A somewhat hidden
assumption is that the accreting plasma is a two-temperature plasma
with proton temperature much greater than the electron temperature.
Only then will $f \simeq 1$.  This can be true only if most
of the ``viscous'' dissipation goes into the protons and there is
no collective effect that efficiently couples the protons and 
electrons.  No such effect has yet been convincingly demonstrated
to exist.  

The assumption that $f$ is constant with radius is only likely to hold
true if $f \simeq 1$ throughout the flow.  We have nonetheless
calculated solutions for constant $f < 1$ in order to illustrate the
effects of a smaller value of $f$ on the dynamical aspects of the
flow.  In the future, it is clear that models will need to include
cooling processes and calculate $f$ self-consistently.  This task is
complicated by the importance of Compton cooling in these flows, which
depends not only on the local conditions at a particular radius, but
also on the incident photon flux from all other radii.  Thus far,
cooling has been included in some detail in some models, but no model
has included both a full treatment of cooling and fully relativistic
dynamics.

Some additional assumptions and limitations of our solutions are tied
to our treatment of angular momentum transport.  For example, we have
vertically averaged the flow.  This should produce a reasonably
reliable solution close to the midplane, but is not predictive for flow
near the poles.  We could produce a full axisymmetric steady-state
solution, but because it is not known how turbulent angular momentum
transport varies with height in the accretion flow, such an effort
would not significantly improve the reliability of the solution.  In
addition, the flow has been assumed smooth and steady.  This is likely
to be true only in a time-averaged sense.  In particular, if $\alpha
\sim 1$, then the turbulence that transports angular momentum will be
only marginally subsonic and shocks and substantial density variations
are likely.

\section{Summary}

We have presented steady-state solutions in the Kerr metric for
advection-dominated accretion onto black holes.  The solutions extend
from $2 \times 10^4 GM/c^2$ down to just outside the event horizon at
$(1 + \sqrt{1-a^2}) GM/c^2$.  The flow passes through a sonic point
and a viscous point which arises due to our use of a causal viscosity
prescription.  We find large variations in the character of the flow
by varying the solution parameters, which include the black hole spin
$a$, the viscosity parameter $\alpha$, the advected fraction $f$, and
the adiabatic index $\gamma$.  We find smooth solutions without shocks
throughout the parameter space.

The innermost part of the flow near the horizon becomes much hotter
and denser as $a$ approaches $1$.  Also, the amount of angular momentum
transferred to the black hole, which is an eigenvalue of our
solutions, drops as $a$ increases.  The structure of the flows is
sensitive to $\alpha$, the efficiency of angular momentum transport.  
For $\alpha ~\sles~ 10^{-3}$, the flows exhibit a pressure maximum and 
an approximately constant specific angular momentum profile, similar
to the ``thick disk'' or ``ion tori'' models.  The same holds true for
solutions with small values of $f$, even though the low-$f$ solutions
are geometrically thin and are similar to the classical thin disk
solutions in a number of respects.  

We find that an equilibrium spin rate $a_{eq}$ exists for black holes
accreting from an advection-dominated flow.  This results from the
decrease in the angular momentum accretion rate as $a$ increases.  At
$a_{eq}$, the accreted angular momentum is just sufficient to
counteract the effects of mass accretion and keep $a$ constant.  For
large values of $\alpha$ and $f$, $a_{eq} \simeq 0.8-0.9$, well below
the maximum value of $a \simeq 0.998$ allowed by photon transport.

\acknowledgments

We thank R. Narayan for helping to initiate this project and for his
support, insight and encouragement, and H. Falcke, J.-P.  Lasota,
M. Rees, and M. Abramowicz for helpful discussions.  This work was
supported by grants NASA NAG 5-2837 and NSF AST 9423209.

\clearpage

\figcaption{Advection-dominated accretion flows for various values of
$a$, all with $\alpha=0.1$, $f=1$, and $\gamma = 1.4444$.  The
solutions shown have $a=-0.999$, -0.9, -0.5, 0, 0.5, 0.9, and 0.999.
The six panels show the radial velocity $V$, density $\rho$, angular
velocity $\W$, specific angular momentum $l \eta$, temperature $T$,
and vertical scale height as a fraction of radius $H/R$.  On the panel
showing $V$, the position of the sonic point is denoted by a filled
circle, and that of the viscous point by a filled square.  On the
panel showing $\W$, the value of $\omega$ at the horizon is given by a
dashed line.}

\figcaption{Similar to Fig. 1, but for flows with $\alpha = 0.001$,
0.003, 0.01, 0.03, 0.1, and 0.3, all with $a=0$, $f=1$, and $\gamma =
1.4444$.}

\figcaption{Similar to Fig. 1, but for flows with $f=0.01$, 0.03, 0.1,
0.3, and 1, all with $a=0$, $\alpha=0.1$, and $\gamma = 1.4444$.}

\figcaption{Similar to Fig. 1, but for flows with $\gamma = 1.3333$,
1.4444, 1.55, and 1.66, all with $a=0$, $\alpha = 0.1$, and $f=1$.}

\figcaption{(a) The angular momentum accretion rate $j$ as a function of
the black hole spin $a$, for $\alpha = 0.001$, 0.003, 0.01, 0.03, 0.1,
and 0.3, all with $f=1$ and $\gamma=1.4444$. The dashed line denotes
$j=2a$; for solutions with $j > 2a$, accretion spins up the black hole
to larger $a$.  (b) same as (a), but for $f=0.01$, 0.03, 0.1, 0.3, and
1, all with $\alpha = 0.1$ and $\gamma = 1.4444$.}

\figcaption{The equilibrium value $a_{eq}$, where $j=2a$, as a function
of $\alpha$ (top panels) and $f$ (bottom panels).  The left panel in
each case shows $a_{eq}$, while the right panel shows $\log
(1-a_{eq})$.} 

\figcaption{
The light lines show isodensity contours in the $\varpi = r
\sin\theta$, $z = r\cos\theta$ plane, where $r,\theta$ are the usual
Boyer-Lindquist coordinates.  Solutions are shown for $\alpha = 0.1,
10^{-3}$ and $a = 0, 0.99$.  The heavy solid lines show the event
horizon (inner curve) and the boundary of the ergosphere (outer
curve).}

\figcaption{The Bernoulli parameter $Be = \eta \sE - 1$ as a function of
radius for the four sets of solutions shown in Figs. 1-4.  The four
panels show solutions for various values of (a) $a$, (b) $\alpha$, (c)
$f$, and (d) $\gamma$.}

\figcaption{The outward viscous energy flux, $-t_t^r$, for
solutions with $a = -0.999$, -0.9, -0.5, 0, 0.5, 0.9, 0.999 and otherwise
standard parameters.  Notice that for $a \gtrsim 0.7$ there is an
outward viscous energy flux at the horizon.}

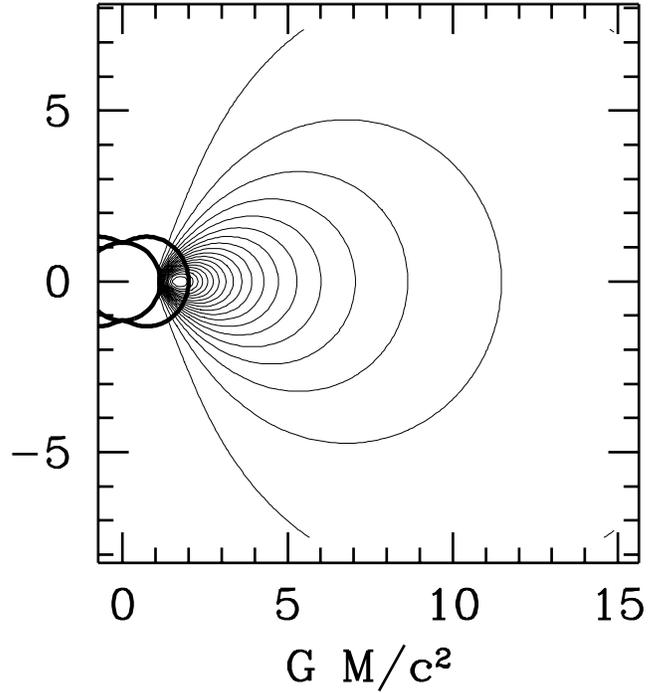
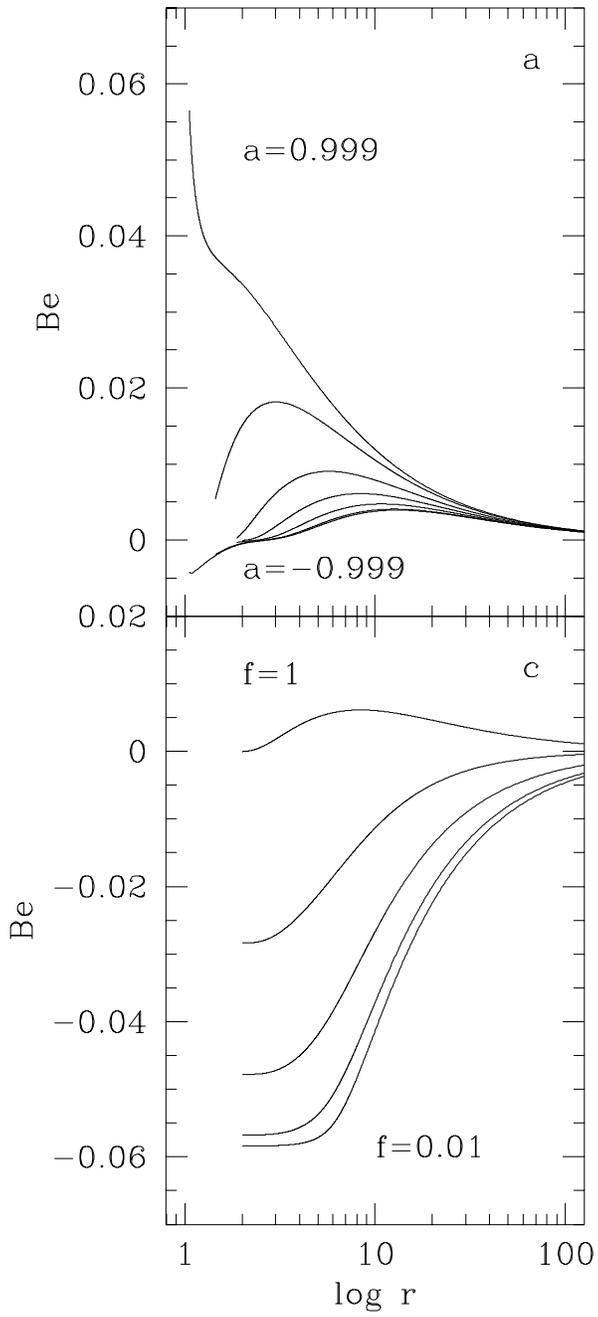
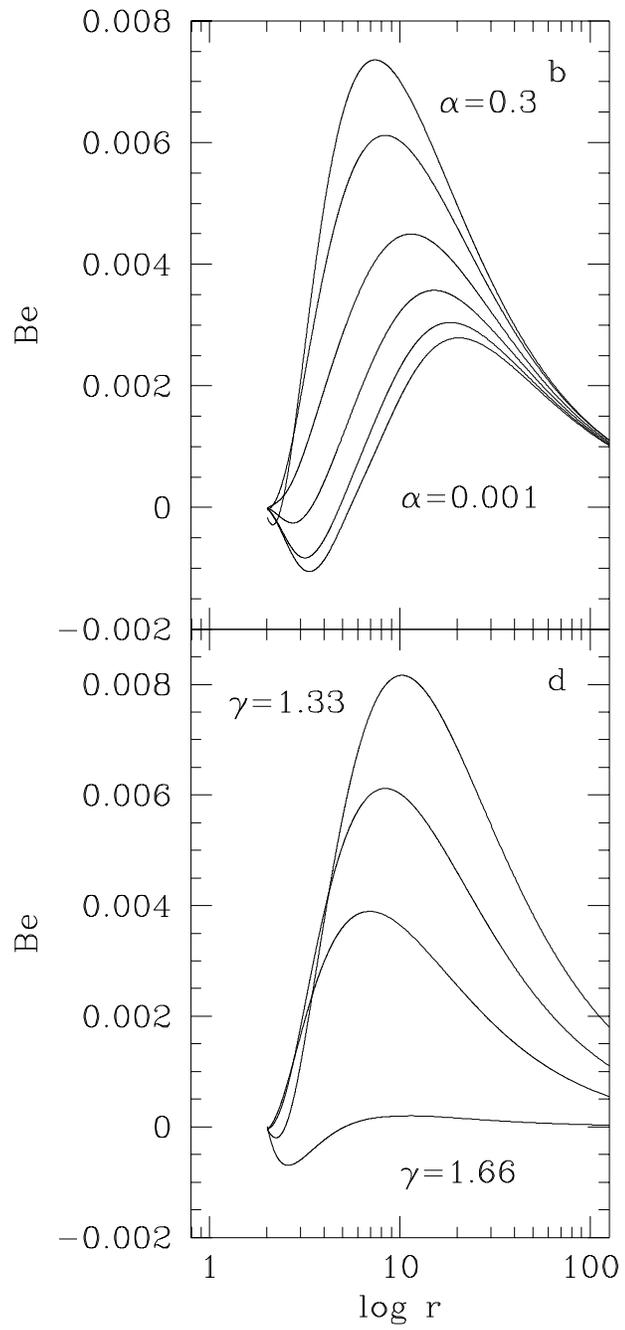
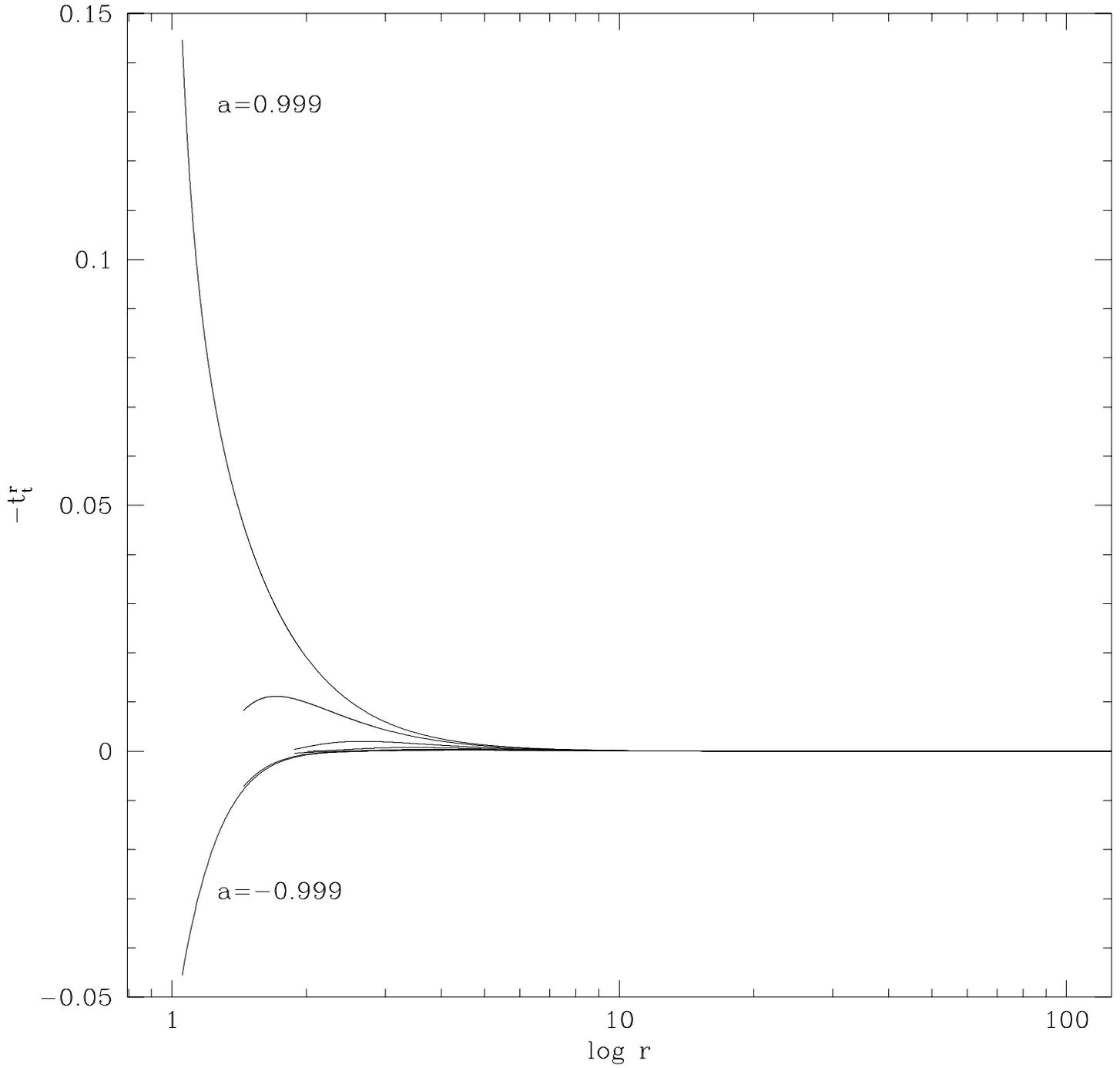
\end{document}